\documentclass[twoside,reqno,12pt]{amsart}
\usepackage{amssymb}
\usepackage{enumerate}
\usepackage{a4wide}
\usepackage[english]{babel}
\usepackage{amsmath}
\usepackage{graphicx}

\newcommand{\beq}{\begin{eqnarray}}
\newcommand{\eeq}{\end{eqnarray}}

\newcommand{\beqq}{\begin{eqnarray*}}
\newcommand{\eeqq}{\end{eqnarray*}}

\newcommand{\bl}{\begin{lemma}}
\newcommand{\el}{\end{lemma}}

\newcommand{\br}{\begin{remark}}
\newcommand{\er}{\end{remark}}

\newcommand{\bex}{\begin{examples}}
\newcommand{\eex}{\end{examples}}

\newcommand{\bt}{\begin{theorem}}
\newcommand{\et}{\end{theorem}}

\newcommand{\bd}{\begin{definition}}
\newcommand{\ed}{\end{definition}}

\newcommand{\bp}{\begin{proposition}}
\newcommand{\ep}{\end{proposition}}

\newcommand{\bc}{\begin{corollary}}
\newcommand{\ec}{\end{corollary}}

\newcommand{\bpr}{\begin{proofs}}
\newcommand{\epr}{\end{proofs}}

\newcommand{\bprm}{\begin{proofsm}}
\newcommand{\eprm}{\end{proofsm}}

\newcommand{\bi}{\begin{itemize}}
\newcommand{\ei}{\end{itemize}}

\newcommand{\ben}{\begin{enumerate}}
\newcommand{\een}{\end{enumerate}}

\newcommand{\R}{\mathbb R}

\numberwithin{equation}{section}
\newcommand{\eps}{\varepsilon}

\newcommand{\la}{\lambda}
\newcommand{\La}{\Lambda}

\begin{document}
\title [Competing HIV Strains  and Immune System Response]
{Competing HIV Strains \\ and Immune System Response}
\author{Thierry Gobron$^1$}
\address{$^1$LPTM, CNRS UMR 8089\\
Universit\'e de Cergy-Pontoise \\ Cergy-Pontoise, France}
\author{Mario Santoro$^2$}
\address{$^2$Dipartimento di Matematica\\
Universit\`a di Roma Tor Vergata\\
Roma  Italia}
\author{Livio Triolo$^2$}
\email{gobron@ptm.u-cergy.fr}
\email{santoro@mat.uniroma2.it}
 \email{triolo@mat.uniroma2.it}
\thanks{}
\begin{abstract}
We consider a simple deterministic model which describes an asymmetric
competition between an immune system with a specific and powerful response,
and a virus with a broad toxicity and fast mutations.
Interest in this model relies on the fact that in spite of it simplicity, it reproduces some of the features
of the asymptomatic phase of the infection by HIV-1. In particular, there is a domain of parameters
in which the dynamics is characterized by the apparition of ``blips'', associated here to an instability
which develops at high virus reproduction rate.
Various possible extensions of this simple model are discussed, in particular in view of its applications
in the context of  HAART therapy.
\end{abstract}
\maketitle \setcounter{equation}{0}

\section { Introduction}
\label{sec:1}

The Human Immunodeficiency Virus (HIV) infects humans in a peculiar way:
it attacks and develops  precisely on the cells which are the keystone of the immune defense system.
A number of theoretical papers have studied the subtle mechanisms
which may describe this evolution and much work in modeling has been
produced since the first attempts by Nowak-May \cite {NM},
Perelson-Nelson \cite {PN} and Nowak-Bangham \cite {NB}.
In spite of true successes, mathematical modeling of such biological systems remains a challenge,
not because of lack of good theoretical approaches, but on the contrary because of intrinsic difficulties
to distinguish experimentally between them.
In this work, we try to escape this kind of dilemma in the following way:  we elaborate on previously considered
deterministic models, but
we consider it from the point of view of strategy, retaining only the following
facts from the original problem: the immune system has a powerful specific response while the virus is
in some sense weaker, but faster to adapt. In other words, we construct a ``predator--predator'' model (rather than a ``predator-prey''),
where the asymmetry comes from the very different strategy
between both.
This results in a model with few parameters, but
a surprisingly rich and suggestive dynamics. Values of parameters, or at least orders of magnitude, are estimated from clinical data, in order
to get some experimental feedback.
Such a scheme seems to us to be new: one one hand, only predator-prey models (in a sense or another) have been considered so far,
and in the other hand, models considering viral mutation don't include it as a part of a deterministic process, but
through an external or stochastic mechanism (see for instance \cite{K}).

One of the main results of the present approach is the appearance of an instability in a sensible range of parameters,
which is strongly reminding of the so-called ``blips''. If it were so, these ``blips'' could be interpreted as interesting probes of the viral dynamics,
rather than uncorrelated, random events.

A number of important features are of course lacking in this model. Some of them (naive cell production, virus fitness landscape, specific behavior under HAART therapy, $\cdots$)
could be rather easily inserted in this frame \cite{PW}, \cite{YSKAC}.  However we believe that the simplicity and robustness of this model is its main strength, and that it can be turned out in an accurate tool without much efforts.

In section \ref{sec:2}, we describe the model, give some hints on its construction and the connection of parameters with clinical data.
In section \ref{sec:3}, we explore its dynamics, and in last section \ref{sec:4}, we discuss some of the features and anticipate on further developments.

\section{Model: scope, construction and parameters.}
\label{sec:2}

In this section , we describe a system of differential equations in which we try to insert as much as possible of the main features
of the HIV-1 infection. Departing from previous modeling based on a predator-prey model in various forms \cite{NM},\cite{K},
we consider the infective virus and the immune system as two predators which have developed different strategies to win over the other:
 the former has a relatively slow dynamics but a strong specific activity while the latter has a fast dynamics, a smaller but largely aspecific toxicity and a high mutation rate.

 First, we define a simple genomic space as a set of indices $S=\{1,\cdots,N\}$
 where $N$ is very large and represents the number of possible virus
 mutants as seen from the immune system. Hereafter we will make very simple assumptions about this space, but both a more realistic topology of the genomic space and a fitness landscape
 could enter here. The viral system is then described as a collection of variables  $\{V_{\sigma}(t)\}_{\sigma\in S}$, giving the density of each
 virus strain at time $t$. The immune system is described as a collection of densities of cells $\{X_{\sigma}(t)\}_{\sigma\in S}$, each specific to a given strain.
 The connection with clinical data is made by identifying the $\{X_\sigma\}$ to the densities of  CD4$^+$ T-helper cells. This choice reflects the fact that 
 this is the limiting factor from the kinetic point of view, effective destruction of viruses taking place in a short time after their
binding to a CD4$^+$ cell. In the same spirit, we do not describe directly a population of infected cells, assuming that the virus dynamics is
fast enough so that both population densities stay strongly coupled at all times. Finally we also assume that the mean densities are the only variable which matters, irrespective of the repartition in the body and the possible interaction with other agents. 

The evolution of these two sets of population densities starting from given initial conditions
 is defined by the following differential equations for all $\sigma \in S$:

\beq\label{Paris1-a}
 \frac{ d X_\sigma}{ d t} &=& \frac{\La_0}{N} \bigl(1 + \frac{D V_\sigma}{1+ V^T} \bigr) - X_\sigma \bigl(1 +  V^T + A V_\sigma\bigr)  \\
  \frac{d V_\sigma}{d t}  &=& \omega V_\sigma \bigl( X^T  -B X_\sigma - C \bigr)+ \eps \left(\Delta V \right) _\sigma
  \label{Paris1-b}
  \eeq
where $X^T$ and $V^T$ are the total population densities for the immune system and the viruses:
\beq\nonumber
X^T = \sum_{\sigma \in S} X_\sigma \qquad V^T = \sum_{\sigma \in S} V_\sigma
\eeq

One of the important simplification we have made here is to avoid the description of a population of naive cells, specializing after (direct or indirect) contact with a particular virus strain. Instead, we distribute initially the (naive) CD4$^+$ cells uniformly over (a segment of ) the
genomic space. This is certainly not the true behavior of the immune system, but has to be considered as a (good) mathematical trick.
We will discuss at the end some of the consequences of this assumption at the mathematical level and what one can expect after removing it.
Here $\Lambda_0$ is the density of  (naive) CD4$^+$ cells in the absence of infection.

For $V^T=0$, they are arbitrarily distributed over
the whole genomic space to avoid the description of an activation process. In the presence of a given strain of viruses, the density of corresponding T-cells increases linearly for low viral load, and is bounded. Here again the choice of this particular dependency
on the virus density is dictated by simplicity. It is however important that it stays bounded at large virus load.
Death rate is made of three parts: a natural death rate, an aspecific
viral death rate proportional to the total virus density, and a specific viral death rate emphasizing an higher probability for a T-cell to be
infected by the viral strain it recognizes.

Viruses replicate through cell infection, so both virus replication and death rates are proportional to strain density. Thus the right-hand side of equation \eqref{Paris1-b} contains three terms proportional to $V_\sigma$: an aspecific replication rate; a term representing the balance between specific replication rate and specific death rate, the former
being larger, thus with an overall negative sign; an aspecific death rate. The last term in  \eqref{Paris1-b} describes the virus mutations. It is taken as a discrete laplacian derived from the metrics ascribed to the genomic space. Various choices for this metrics are possible.  In the following, we will always assume that all strains have the same fitness, so that the metrics depend on the connectivity only.

In spite of its simplicity, this model is thought to be able to capture some of the features of the specific immune response to HIV infection. The metrics being given, the model described above depends on $6$ parameters: $A$, $B$, $C$, $D$, $\omega$ and $\eps$. Three additional scaling parameters fix the density and time units.

In order to get an expression for them in terms of clinical data, we first re-write the evolution equations for the ``true'' quantities:
$\tilde{X}_{\sigma}$, densities  of CD4$^+$ cells,  $\tilde{X}_{\sigma,\tau}^*$, densities  of cells of type $\sigma$ infected by a virus of type $\tau$, $\tilde{V}_\sigma^*$, densities of viruses and $\tilde{t}$,  time in days. We write

\beq\label{Paris1a}
 \frac{ d \tilde{X}_\sigma}{ d \tilde{t}} &=& \frac{\la_0}{N} \bigl(1 + \frac{d  \tilde{V}_\sigma}{d_X+ k  \tilde{V}^T} \bigr)
  -  \tilde{X}_\sigma \bigl(d_X +  k  \tilde{V}^T + k'  \tilde{V}_\sigma\bigr)  \nonumber\\
   \frac{ d \tilde{X}_{\sigma,\tau}^*}{ d \tilde{t}} &=&
  \tilde{X}_\sigma \bigl(  k  \tilde{V}_\tau + k'  \delta_{\sigma,\tau} \tilde{V}_\sigma \bigr)
  - d_{X^*}   \tilde{X}_{\sigma,\tau}^* -\mu  \tilde{X}_{\sigma,\tau}^* \tilde{X}_\tau
  + \rho \Delta _\tau\left( \tilde{X}_{\sigma,\tau}^* \right)
  \\
   \frac{d \tilde{V}_\tau}{d \tilde{t}}  &=& Z d_{X^*} \sum_{\sigma}  \tilde{X}_{\sigma,\tau}^*
  - d_V  \tilde{V}_\tau
   \nonumber
\eeq

Using data from \cite{PN},\cite{RP} and references therein, estimates for coefficients are:
{\obeylines
$\la_0= 10^4$ ml$^{-1}$ day$^{-1}$
$d_X = 0.01$ day$^{-1}$
$k= 2.4 10^{-8}$ ml day$^{-1}$
$d_{X^*} = 1. $ day$^{-1}$
$Z= 10^3 -10^5$ virions per infected cell
$d_V=23$ day$^{-1}$
$\rho = 10^{-5}$ day$^{-1}$
}
\vspace{6pt}
Other parameters are less easy to estimate. $d$ represents the capacity of the immune system to reinforce its specific response;  $k' > 0$ indicates that CD4$^+$ cells have a larger probability to be infected by their specific virus strain. Both are believed to be positive and reasonably larger than  $k$; $\mu$ should be large enough
to allow for an effective specific immune response. Assuming a fast dynamics for the viruses, we set $ \frac{d \tilde{V}_\sigma}{d \tilde{t}}=0$ at all time, which induces a relationship between infected cells and virus densities:

\beq
 \tilde{V}_\tau = \frac{Z d_{X^*}}{d_V} \sum_{\sigma}  \tilde{X}_{\sigma,\tau}^*
\eeq

Inserting this expression in the equation for the infected cells leads to an equation for the virus densities on a slow time scale.
Under scaling $t =  d_X \tilde{t}$,  $ X_\sigma =  \frac{d_X}{\lambda_0}\tilde{X}_\sigma$, $V_\sigma = \frac{k}{d_X}  \tilde{V}_\sigma $, the system of equations \eqref{Paris1-a}--\eqref{Paris1-b} is recovered with $\Lambda_0 = 1$ and the following expressions and orders of magnitude for the other parameters:
{\obeylines
$A=\frac{k'}{k}$
$B=\frac{\mu\,d_V}{k\, Z\, d_{X^*} } -\frac{k'}{k} \approx 10^{6}$ (with $\mu = 1$ ml  day$^{-1}$ and $Z= 10^3$).
$C=\frac{d_X\, d_V}{Z \,k \,d_{X^*}}\approx 1$ (with $Z= 10^3$).
$D= \frac{d}{k} $
$\omega=\frac{k \, Z \,d_{X^* \, \lambda_0}  }{d_V\,d_X^2}\approx 10^{2}$
$\eps = \frac{\rho}{d_X}\approx 10^{-3}$
}
The values for $A$ and $D$ are not known but will be taken in the following in the range $[1,10^2]$. Note that the scaling factors of CD4$^+$ cells and virions are of the same order of magnitude so that in the reduced coordinates a density of order $1$ corresponds to a true density of about $10^6 \mu l^{-1}$, while the reduced time unit is of order $100$ days.

In the following section, we will explore the behavior of this model, first by considering a reduced model with no mutations, which behavior will help then to analyze the full model.

\section{Dynamics}
\label{sec:3}

\subsection{Stationary solutions in the absence of mutations}\hfill\break

In the absence of mutations, it is not hard to show that the set of stationary solutions is particularly simple: in terms of viral populations, either there is no virus, $V_\sigma=0$
for all $\sigma\in S$, or there exist $K$ strains of virus of equal density, $1\le K\le N$. In this subsection, we analyse the behavior of these solutions,
for two reasons: first, it should give some insights on the full system when mutation rates are small with respect to other parameters; second, it gives a connection with previously studied models \cite{NM} where mutations are not explicitly  taken into account.

The simplest stationary solution is the configuration with no virus, $V^T=0$, $X^T=\Lambda_0$, which exists for all values of parameters.
It is locally stable for $C>\Lambda_0 (1-\frac{B}{N})$ and locally unstable otherwise.

The other stationary solutions can be described as follows. Let $K$, $1\le K\le N$,  be the number of virus strains with positive density in a given stationary solution with viruses. The densities of these strains are necessarily equal and the overall densities of  CD4$^+$ cells and viruses are related by
\beq\label{xt_eq}
X^T = \frac{\Lambda_0}{1 + V^T} \bigl( 1 +\frac{K}{N}\frac{(D-A) V^T}{K +(K+A) V^T}\bigr)
\eeq

while the density of viruses $V^T$ is solution of the following equation
\beq\label{c_eq}
C= \frac{\Lambda_0}{N} \bigl( \frac{A(N-K)-D(B-K)}{A(1+V^T)} +  \frac{K (B-K)(D-A)}{K +(K+A) V^T}\bigr)
\eeq
This equation has at most two real positive solutions in $V^T$, depending on the value of the parameters. The role of the parameters as well as a qualitative behavior of these solutions can be drawn directly from the above two formulas:
\begin{figure}[htbt]
\begin{center}
\includegraphics[scale=.5,angle=-90]{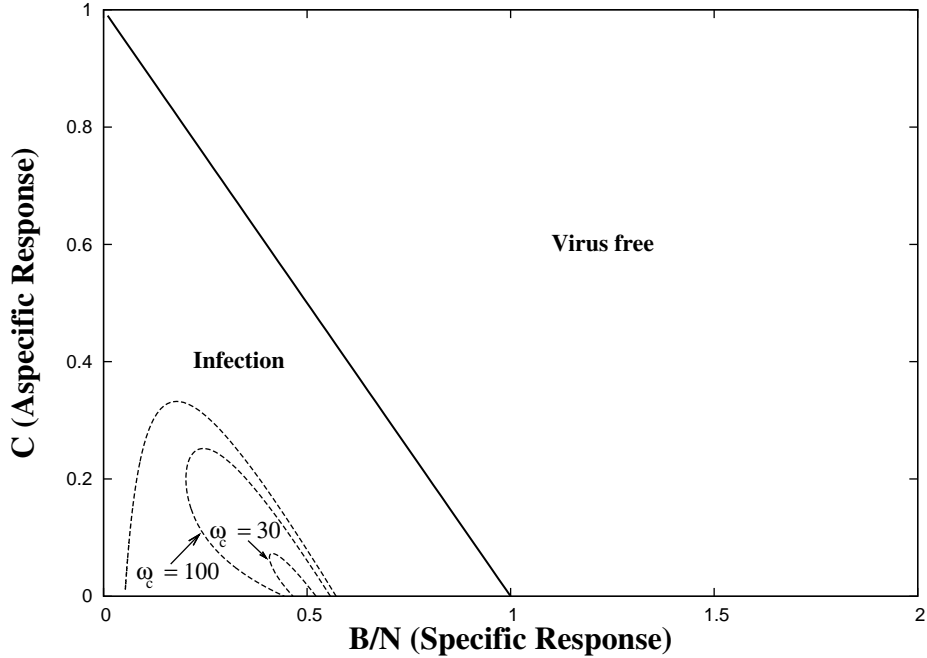}
\end{center}
\caption{Domain of existence of solutions in the plane $(B,C)$ for $D >A$ ($D=30$, $A=10$). The linear stability limit of the virus free solution $C=\Lambda_0 (1-\frac{B}{N})$ sets the limit of existence of solutions with $V^T\not= 0$. Dashed lines: boundaries of domains of existence of limit cycles for $\omega= 30$, $\omega= 100$, $\omega= +\infty$ (for $K=5$).}
\label{figure:existence1}
\end{figure}
\begin{figure}[htbt]
\begin{center}
\includegraphics[scale=.5,angle=-90]{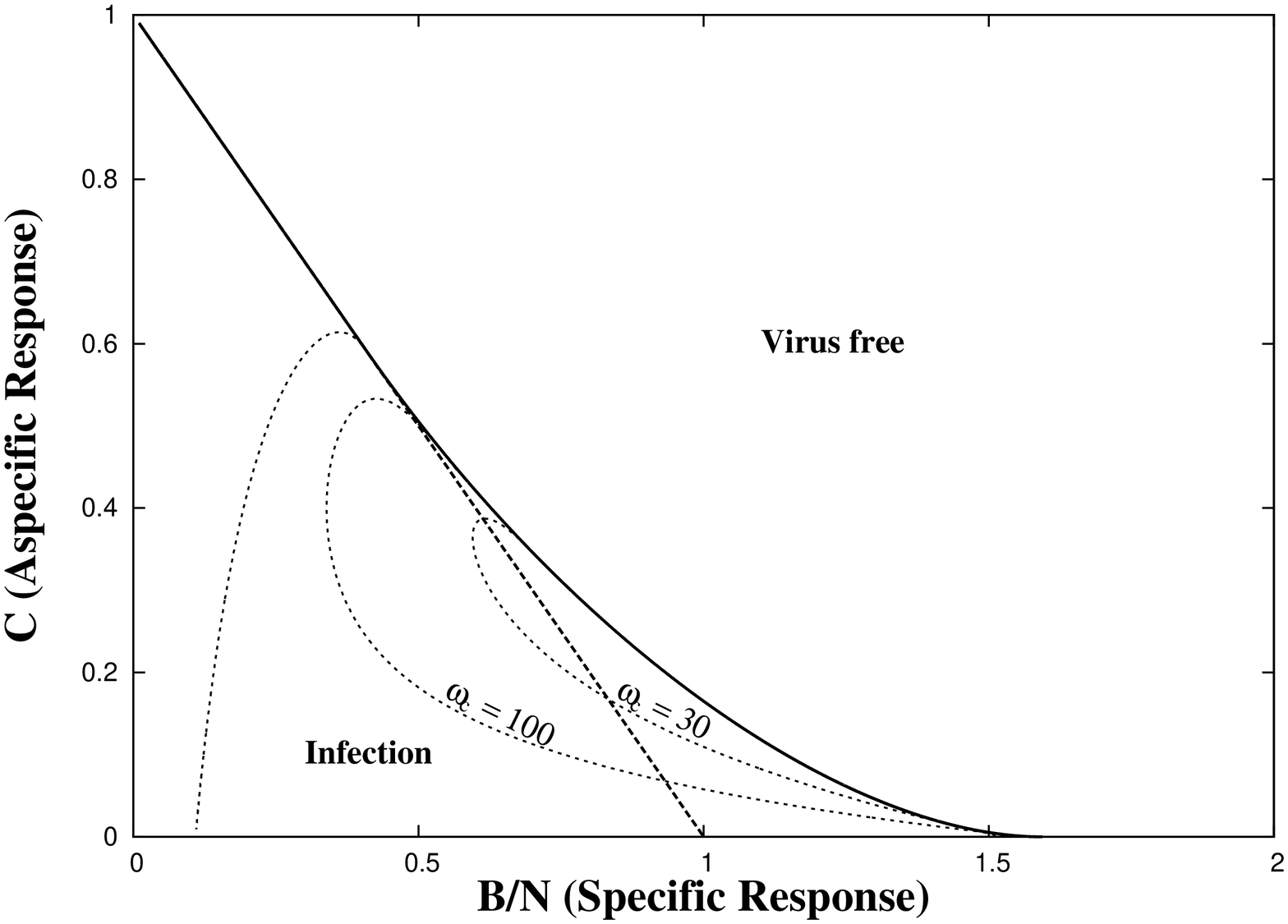}
\end{center}
\caption{Domain of existence of solutions in the plane $(B,C)$ for $D <A$. This domains extends now beyond the linear stability limit of the virus free solution $C=\Lambda_0 (1-\frac{B}{N})$. Here $D=15$, $A=30$. Dashed lines: boundaries of domains  of existence of limit cycles  for $\omega= 30$, $\omega= 100$, $\omega= +\infty$ (for $K=10$).}
\label{figure:existence2}
\end{figure}

The value of $C$ reflects both the ``bare'' virus death rate and the a-specific response of the immune system which is not described explicitly
in this model. Since the right hand side of \eqref{c_eq} is bounded by $\Lambda_0$, the virus density will be zero in the stationary regime if $C$ is high enough, irrespective of the behavior of the specific immune system.

The specific response is described by two parameters, $B$, the specific cytotoxicity,  and $D$ which controls the increase of specific CD4$^+$ cell production in presence of the corresponding virus strain. The action of the virus is controlled by $\omega$, the virus reaction rate scaling factor, $N$ the size of the system on which the virus lives,  and $A$ the killing rate of specific CD4$^+$ cells.  The values of the ratios $\frac{B}{N}$ and $\frac{D}{A}$ control the behavior of equilibrium solutions in two different ways:

When $N > B$ and  $C <  \frac{\Lambda_0 (N-B)}{N} $,  the values of $C$ and $B$ are too small to eradicate the virus and there is one solution for each $K\ge 1$. If $\frac{D}{A} > \frac{N}{B} > 1$ the production of specific CD4$^+$ cells is high enough to limit the density of viruses ($V^T$ is bounded in the limit $C\longrightarrow 0$), provided
\beq\label{K_lim}
K< \frac{D B - A N}{(N-B) + (D-A)}
\eeq
For larger values of $K$ or smaller values of $D$, the specific response has a very little effect in the sense that $V^T$ could grow without bounds as $C$ goes to zero.

In the other case,  $C > \frac{\Lambda_0 (N-B)}{N} $, the immune response is high enough to sweep out low densities of viruses. However if in addition $\frac{D}{A} < \frac{N}{B}$, a sufficiently high densities of viruses
will deplete the specific  CD4$^+$ cell population and a pair of solutions with a small number of high density virus strains may appear (for $C$ low enough and $K$ as in \eqref{K_lim}). For   $\frac{D}{A} > \frac{N}{B}$ this density effect is too small and the unique solution will be virus free.

The linear stability of the stationary solutions with $V^T\not=0$ reveals interesting behaviors. We first recall that whenever there are two stationary solutions for fixed $K$, the one with the smallest density is necessarily unstable and have to be discarded. The first interesting instability is the following: When $D >A$, any new strain of viruses with small positive density will tend to grow out  while for $A >D$, strains with densities slightly smaller than the equilibrium one will die out. This effect may appear in some sense paradoxical, but recalls a known behavior of HIV infection: if the specific toxicity $B$ is too small to eradicate the virus, increasing the population of CD4$^+$ cells does not help and contributes instead to virus growth. This instability will be also important in presence of mutations as the ratio $\frac{D}{A}$ will control the number of strains.

The other possible linear instability is associated with the apparition of a pair of complex eigenvalues of the linearized map, with positive real part.
When the following expression is positive,

\beq
\omega_c^{-1} &=& \frac{\lambda_0 V^T}
{N (1+V^T) (K+(K+A) V^T)}\nonumber\\
&&\qquad\times
\biggl(
\frac{  K  \bigl((B-K) (A-D) - K (N-B)\bigr)}
{ 2 K + (2 K +A) V^T}
+\frac{D (B-K)  V^T}
{(1+V^T)^2 }
\biggr)
\eeq

then  for  all $\omega >\omega_c$, the solution with $K$ strains is unstable. Such an instability  generally indicates the existence of a limit cycle rather than a stable fixed point. We will further explore this point in the next subsection which is devoted to the dynamics in a restricted setting.

\subsection{Non linear evolution of $K$ identical strains}\hfill\break

In this subsection, we analyze the evolution in the absence of virus mutations, starting from a set of particular initial conditions in which $K$ virus strains are present with the same density, and T-cell densities depend only on the presence or absence of the corresponding viral strain. By symmetry, these properties will be preserved by time evolution, and the system will be described by three densities:
Activated T-cells $X_a$, non-activated T-cells $X_n$ and virus $V$. Here $V^T= K V$ and $X^T= K X_a + (N-K) X_n$. The time evolution is given by a system of three coupled differential equations, with an explicit dependence on the number of strains $K$.

\beq
\frac{d X_a}{d t} &=& \frac{\Lambda_0}{N}(K+\frac{D V^T}{1 + V^T} ) -X_a (1+ \frac{K +A}{K} V^T )\nonumber\\
\frac{d X_n}{d t} &=& \frac{\Lambda_0 (N-K)}{N} -X_{n} (1+ V^T )\nonumber\\
\frac{d V^T}{d t} &=&\omega V^T ( X_{n} - \frac{B-K}{K} X_a  -C )
\eeq

\begin{figure}[!htbp]
\begin{center}
\includegraphics[scale=.5,angle=-90]{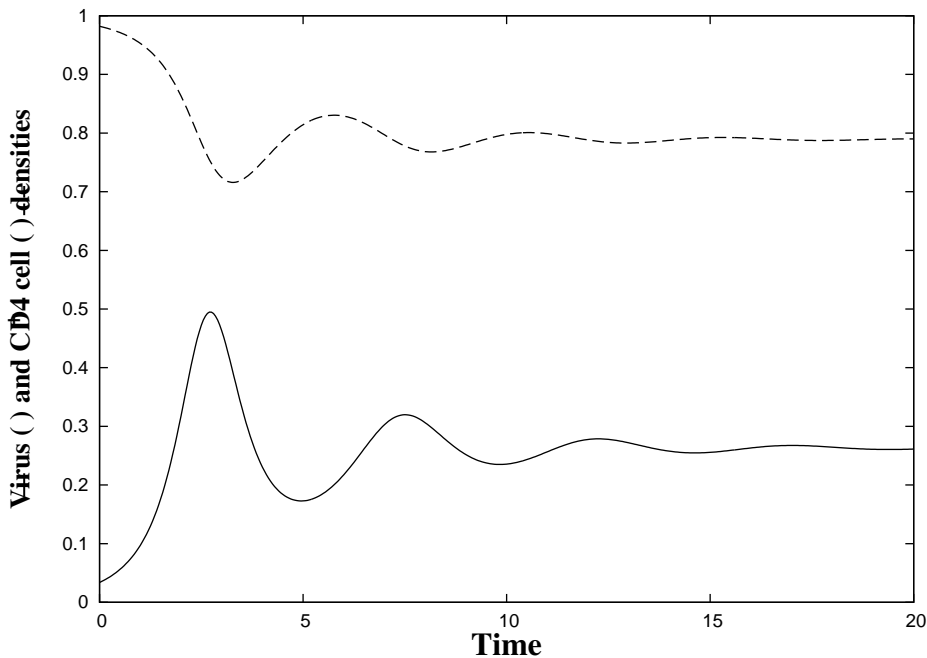}
\end{center}
\caption{
Time evolution of virus (continuous line) and CD4$^+$ (dashed line) densities toward a stable equilibrium point starting from an initially low virus density. Here $K=10$, $\frac{B}{N}=.36$, $C=.6$, $A=56.$, $D=21.$, $\omega=20.$.
}
\label{figure:fixedK-S}
\end{figure}

The solutions $(X_a(t), X_n(t), V(t))$ are bounded in $(\R^+)^3$ at all times provided there are so initially.
There is at most three admissible stationary solutions which coincide to those found in the previous subsection, and the discussion on their existence and stability is the same as above, except for the variation in the number of strains which is here blocked by construction. 
The system has either a stable stationary solution or a limit cycle (in the $(X^T, V^T)$ plane ) and the dependence on the parameters is the same as in figures \eqref{figure:existence1} for $D>A$ and \eqref{figure:existence2} for $D<A$. Numerical simulations are an easy way to get informations on the dynamics. When there exists a stable stationary solution with non zero virus density, a typical behavior consists in an initial virus peak and a successive lower CD4$^+$ density as in figure  \eqref{figure:fixedK-S}. Note that the lowering of CD4$^+$ density is smaller than what is expected from clinical data. This is probably due to the mode of creation of ``naive'' cells in this model which creates an excessive rigidity in the CD4$^+$ density distribution.

\begin{figure}[!bhtp]
\begin{center}
\includegraphics[scale=.5,angle=-90]{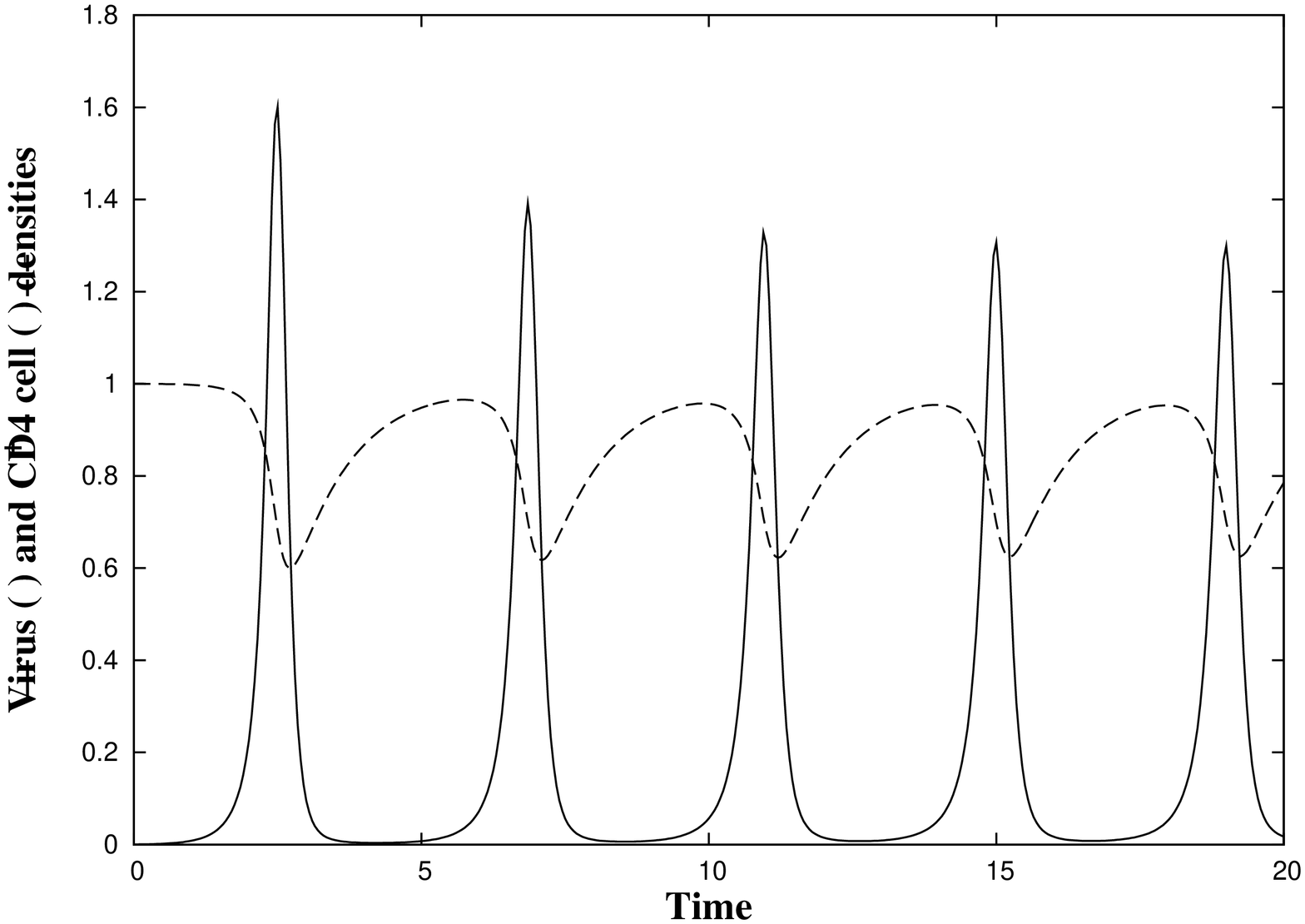}
\end{center}
\caption{
Time evolution of virus (continuous line) and CD4$^+$ (dashed line) densities for larger virus creation rate. Virus density develop sharp peaks reminiscent of so- called ``blips'' . Same parameters as in figure \ref{figure:fixedK-S} with a larger value for $\omega= 70.$
}
\label{figure:fixedK-I}
\end{figure}

When the virus creation rate is high enough, $\omega > \omega_c$, the stationary solution becomes unstable and the system converges to a periodic solution around a limit cycle in the $(X^T, V^T)$ plane, with a period of a few time units (about one year in the original variables). As shown in figure \eqref{figure:fixedK-I} , CD4$^+$ has a slightly oscillating profile, while virus density has very sharp peaks. These peaks are reminding
the so-called blips which may appear in the dynamics of HIV-1 infection. This suggests that ``blips'' are a consequence of a dynamical instability in the virus dynamics rather than a random, uncorrelated, event. If this were the case, the parameters of the blips would be necessarily related
to those of the main dynamical course.

\subsection{Dynamics of mutable virus}\hfill\break

We now consider the full dynamics described by equations \eqref{Paris1-a}--\eqref{Paris1-b}, including mutations which we detail now.  In order to keep with simple hypothesis, we give to our ``genomic space'' a one dimensional topology, thinking of it as a coordinated sequence of favored strains, rather than a full set of possible mutations, which would have to be more complex and should come along with a notion of fitness landscape. Here we consider that all (accessible) strains have the same parameters and can mutate to their ``neighbors'' at a constant rate. Thus we write the mutation term $(\Delta V)_\sigma$ appearing in equation \eqref{Paris1-b}, as a one dimensional discrete laplacian:
\beq
 (\Delta V)_\sigma = V_{\sigma + 1} + V_{\sigma -1} -2 V_\sigma
 \eeq
 We do not explicitly consider boundary conditions as filling up the full system is meaningless. When $D >A$, the system may develop traveling wave solution. Here we consider the opposite case $A>D$ large enough, when competition between strains opposes to constant mutation and possibly leads to a stationary number of strains.  
\begin{figure}[!htbt]
\begin{center}
\includegraphics[scale=.5,angle=-90]{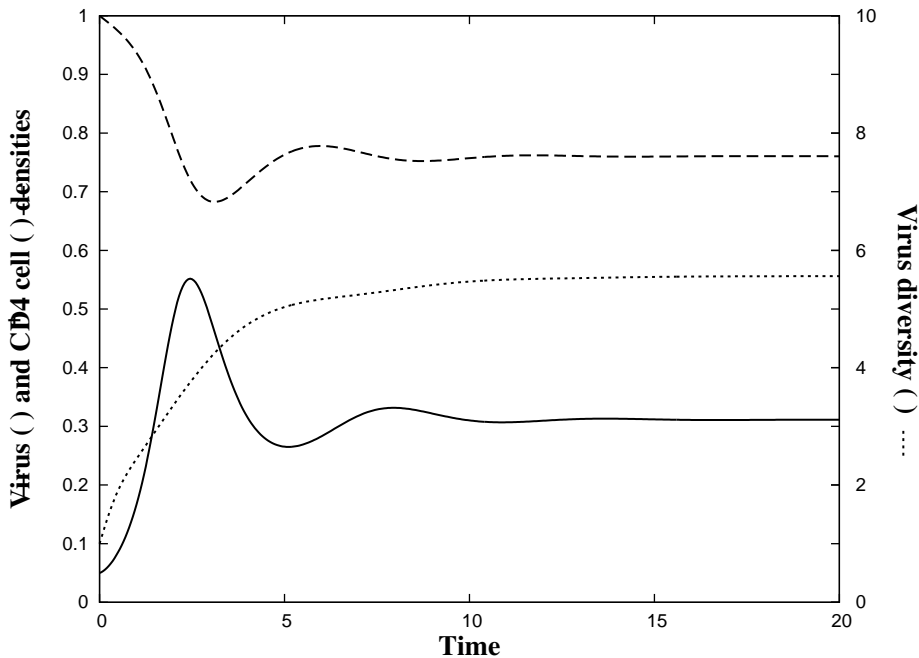}
\end{center}
\caption{
Time evolution towards a stable equilibrium in presence of mutations. The dotted line represents the virus diversity
(see text) and stabilize around $6$. $\frac{B}{N}=.36$, $C=.6$, $A=56.$, $D=21.$, $\omega=10.$, $\epsilon=.5$.
}
\label{figure:Mutation-S}
\end{figure}
We first define the virus ``diversity'' as the quantity $K^*$,
 \beq
 K^* = \frac {(\sum_\sigma V_\sigma)^2}{\sum_\sigma V_\sigma^2}
 \eeq
 This formula is the inverse of the Simpson index \cite{NM} and gives a measure of how many strains are present. When all of these virus strains have exactly the same density as in previous subsections, it is just equal to the number of strains $K$;  for more general distributions, it evaluates the number of strains with a noticeable density.

\begin{figure}[!hbt]
\begin{center}
\includegraphics[scale=.5,angle=-90]{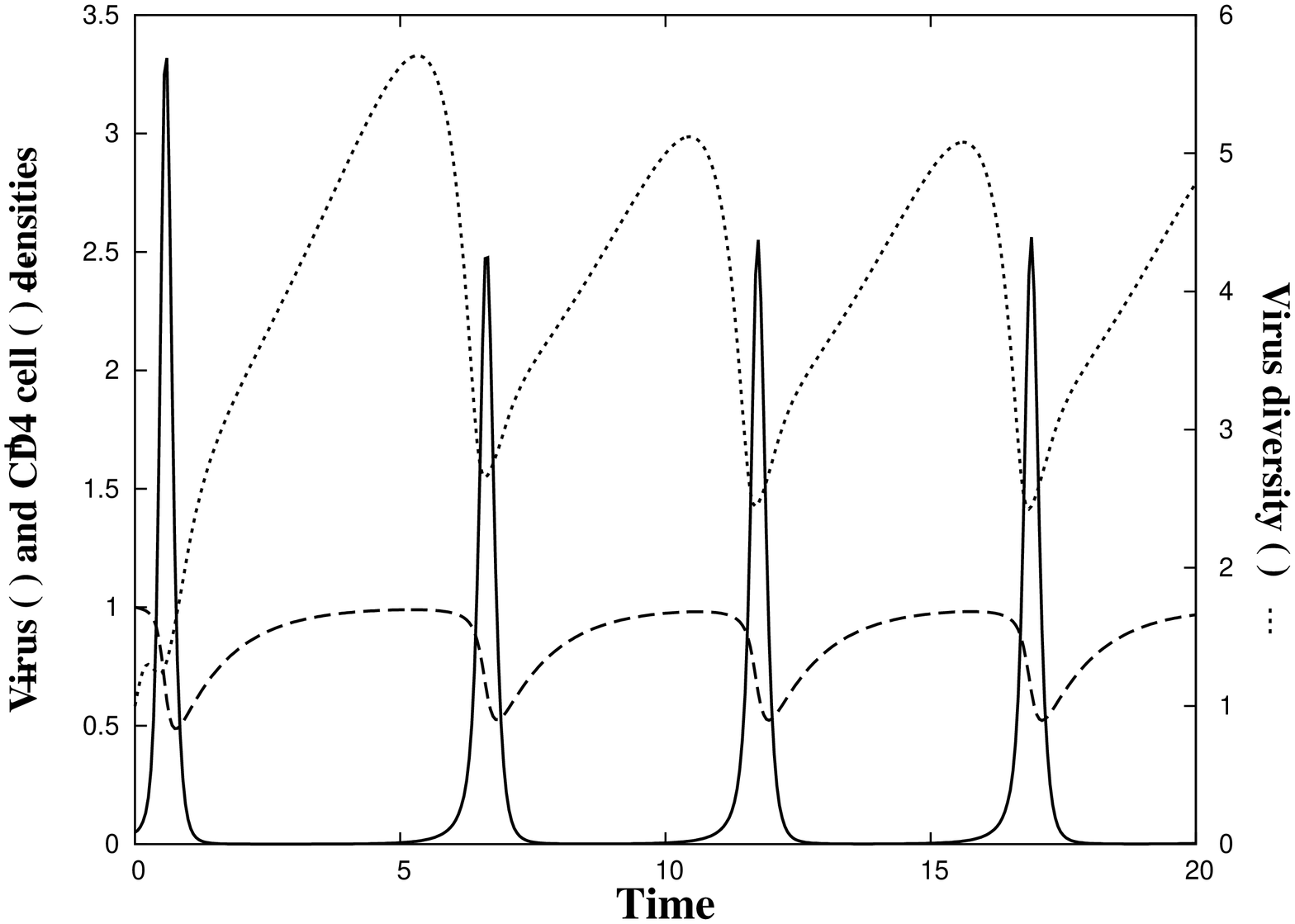}
\end{center}
\caption{
Time evolution towards a limit cycle in presence of mutations. The dotted line represents the virus density  (see text) which oscillates between  $2$ and $5$. $\frac{B}{N}=.36$, $C=.6$, $A=56.$, $D=21.$, $\omega=54.$, $\epsilon=.3$.
}
\label{figure:Mutation-I}
\end{figure}

\begin{figure}[htbt]
\begin{center}
\includegraphics[scale=.5,angle=-90]{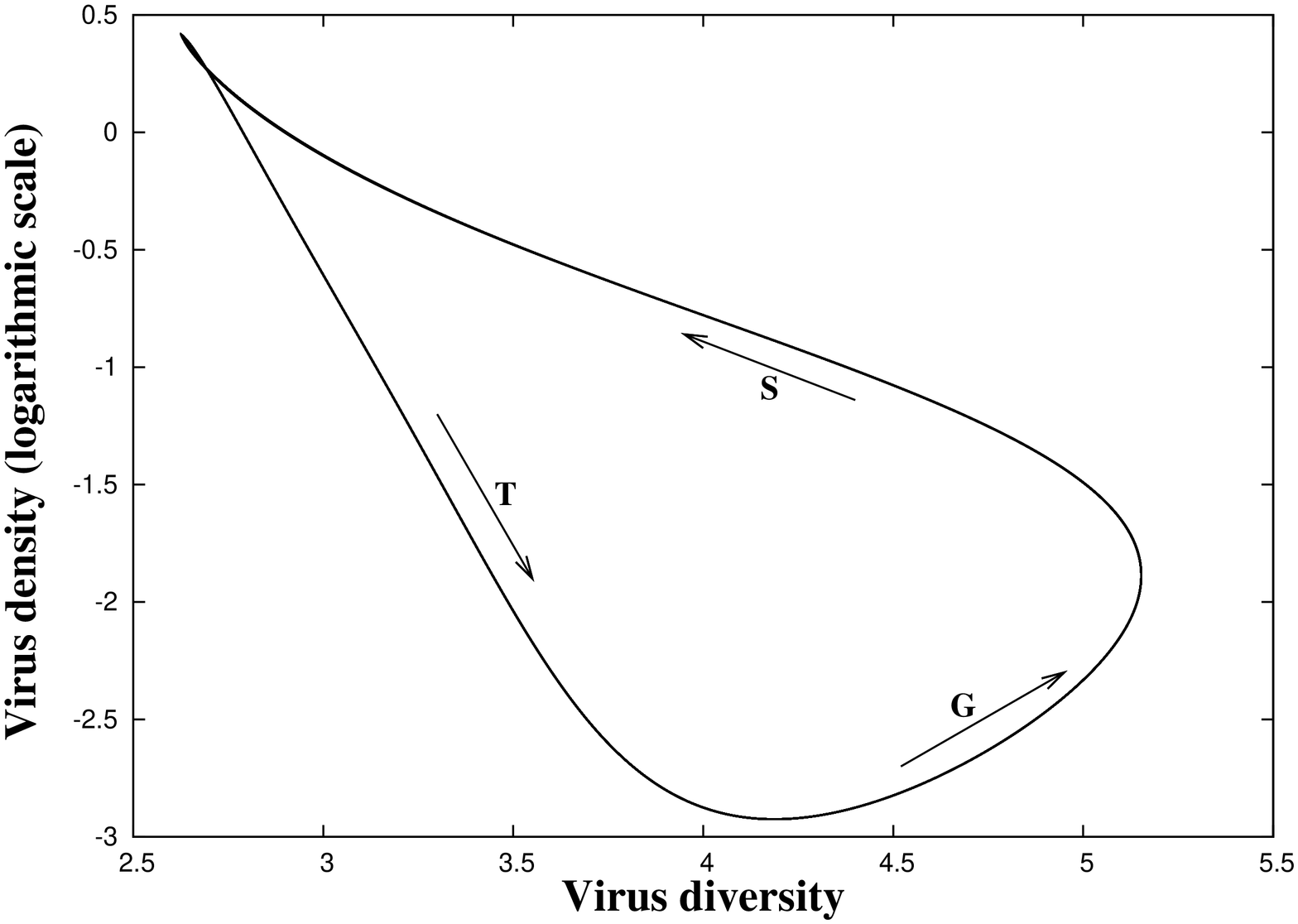}
\end{center}
\caption{
Limit cycle for the virus distribution in the plane diversity-density. The cycle is roughly triangular and each of its side can be associated to
a particular modification of the virus distribution: $G$: initial growth, $S$ selection of strains, $T$ toppling.
$\frac{B}{N}=.36$, $C=.6$, $A=56.$, $D=21.$, $\omega=54.$, $\epsilon=.5$.
}
\label{figure:GST}
\end{figure}

 For values of parameters compatible with the existence of a stationary solution in absence of mutations, one expects that for small vales of the mutation rate, the system has a stationary solution close to the case without mutations. The time evolution now involves the number of strains and a typical example of this behavior is shown in figure  \ref{figure:Mutation-S} where the number of strains grows from initially one to roughly six at equilibrium. The situation is more complicated in the domain of parameters allowing for the presence of blips: now the number of strains oscillates with the virus density, linearly increasing between peaks, and sharply dropping just before the peak. In this latter case,  the limit cycle
has typically a triangular shape in the plane 'number of strains'-'virus density'. This is shown in figure \ref{figure:GST} where each side is characterized by a particular transformation of the virus density profile (G) growth of viral load, (S) Selection of  strains, (T) Toppling of the selected strains and apparition of new ones. Again, this evolution of the virus density profile is strongly reminiscent of so-called ``blips'' \cite{NKK} which are present in infections by HIV-1. Note that in this model, there is no notion of fitness, but there is a selection mechanism. In this pure deterministic model, the index of selected strains depend of course on the initial conditions but any small random perturbation would blur this point.

\section {Conclusion and outlook.}
\label{sec:4}

In this work, we have devised a simple deterministic model, trying to catch some of the main issues of HIV-1 infection from the point of view of strategy, rather than describing as much as possible of this complex process. This model can be viewed as a ``predator-predator'' model in the sense that the birth rates of both increase with the density of the other, and the richness of the dynamics comes from the asymmetry in the behavior of both parts: the virus has fast replication and mutation rates and a broad toxicity while the immune system has a stronger but slowly reacting specific response.

High virus mutability is a key feature dictating most of the structure of the model, while the strength of the specific immune response allows to use a very simple notion of  ``genomic space'' in which mutations can be imbedded, and keep only a very small number of parameters, most of which being related to clinical data. Various kind of time evolutions are found, in which dynamical relations between virus density and virus diversity may show up. In particular, peaks of virus density appear in some cases as a consequence of a too high virus replication rate. Such phenomena is strongly reminding of the clinically observed blips and it is thus tempting to suggest that blips are the consequence of a dynamical instability and not random or unrelated events. Correlations in the clinical data may possibly be found in support to this hypothesis.

A number of important issues have been left aside in this model. The simplest generalization is the introduction of a compartmentalization to mimic the decrease of viral load under efficient HAART therapy. However, the main interest here would be to model virus rebound under therapy and this requires deeper modifications. By construction, our specific immune system can cover efficiently only a small genomic space, and only its interpretation as a sequence of favored viral strains is meaningful. A first step would be an enlargement of this space, the introduction of a less naive topology and a fitness landscape, with possible changes induced by therapy. In turn, maintaining an efficient specific immune response on such a large space requires its confinement essentially where virus is present. And a good way to implement it is the introduction of  a population of naive cells and specialization mechanisms. Such a more refined model is in our view the next step to aim for.

{\bf Acknowledgments}
We thank E. Presutti for scientific discussions on this subject at early stages of this work and C. F. Perno with all his virology team for many useful discussions on biological aspects.
We also acknowledge the financial support of CNRS, Universit\'e de Cergy-Pontoise (France),  Italian PRIN 2007 (coord. Presutti)
 and French-Italian GDRE ``GREFI--MEFI''.

\bibliographystyle{amsalpha}

\end{document}